# Observation of vortex-string chiral modes in metamaterials

Jingwen Ma[1], Ding Jia[3,7], Li Zhang[2,4,5,6], Yi-jun Guan[3,7], Yong Ge[3,7], Hong-xiang Sun[3,7,*], Shou-qi Yuan[3,7], Hongsheng Chen[2,4,5,6], Yihao Yang[2,4,5,6,*], Xiang Zhang[1,*]

[1] Faculties of Science and Engineering, The University of Hong Kong, Hong Kong, China.

[2] Interdisciplinary Center for Quantum Information, State Key Laboratory of Extreme Photonics and Instrumentation, ZJU-Hangzhou Global Scientific and Technological Innovation Center, Zhejiang University, Hangzhou 310027, China.

[3] Research Center of Fluid Machinery Engineering and Technology, School of Physics and Electronic Engineering, Jiangsu University, Zhenjiang 212013, China.

[4] International Joint Innovation Center, The Electromagnetics Academy at Zhejiang University, Zhejiang University, Haining 314400, China

[5] Key Lab. of Advanced Micro/Nano Electronic Devices & Smart Systems of Zhejiang, Jinhua Institute of Zhejiang University, Zhejiang University, Jinhua 321099, China

[6] Shaoxing Institute of Zhejiang University, Zhejiang University, Shaoxing 312000, China

[7] State Key Laboratory of Acoustics, Institute of Acoustics, Chinese Academy of Sciences, Beijing 100190, China.

*Correspondence to: (Y. Y.) yangyihao@zju.edu.cn; (H. S.) jsdxshx@ujs.edu.cn; (X. Z.) president@hku.hk

**As a hypothetical topological defect in the geometry of spacetime, vortex strings play a crucial role in shaping the clusters of galaxies that exist today, and their distinct features can provide observable clues about the early universe's evolution. A key feature of vortex strings is that they can interact with Weyl fermionic modes and support topological chiral-anomaly states with massless dispersions at the core of strings. To date, despite many attempts to detect vortex strings in astrophysics or to emulate them in artificially created systems, observation of these topological vortex-string chiral modes remains experimentally elusive. Here we report the experimental observation of such vortex-string chiral modes using a metamaterial system. This is implemented by inhomogeneous perturbation of a Yang-monopole phononic metamaterial. The measured linear dispersion and modal profiles confirm the existence of topological modes bound to and propagating**

**along the vortex string with the chiral anomaly. Our work not only provides a platform for studying diverse cosmic topological defects in astrophysics but also offers intriguing device applications as topological fibres in signal processing and communication techniques.**

Topology, originally a mathematical subject describing the conserved global quantities of geometric objects, now plays an indispensable role in various aspects of modern physics. In high-energy physics, topological defects, including Weyl, Dirac, and Yang monopoles(*1-3*), are powerful handles to describe various elementary particles. For instance, Dirac fermions describe the behaviour of electrons and quarks in the Standard Model, and Weyl fermions were once thought to describe the neutrinos for a long time until the discovery of neutrino oscillation. Recently, artificially engineered systems such as cold atoms(*4*), condensed matters(*5*), and photonic/phononic metamaterials(*6, 7*) have emerged as versatile platforms for observing the low-energy counterparts of these topological phenomena(*8-11*). Realisations of Weyl and Dirac monopoles in crystalline lattices have unveiled interesting phenomena that are otherwise experimentally unavailable in high-energy physics, such as surface Fermi arcs(*8, 12*), chiral anomaly(*13*), and Dirac hierarchy(*14*). The Dirac monopoles can also be generalised to the so-called Yang monopoles in five dimensions, which have been investigated in cold-atom and photonic systems, leading to direct observations of non-abelian second Chern numbers(*15*) and linked Weyl surfaces/arcs(*16*).

Among many fascinating topological phenomena, vortex strings (or cosmic strings) (*17*) stand out as a typical object that has attracted broad interest in many areas of physics. Vortex strings emerge from the spontaneous symmetry breaking of Higgs field in the early universe and may have acted as seeds for galaxy formation in our cosmos(*18*). Direct observations of vortex strings remain elusive in astrophysics, but similar concepts have been emulated in low-energy systems, including type II superconductors(*19*), liquid helium(*20*), and liquid crystals(*21*). In 1985, Edward Witten made a significant theoretical prediction about vortex strings(*22*): when interacting with Weyl fermionic fields, vortex strings can support chiral-anomaly propagating modes with massless dispersions. Such chiral modes can support the flow of fermions without external electrical fields, making the vortex strings act as "superconducting strings", which might provide observable clues about the evolution of the early universe(*23*). More recently, this key feature of vortex strings — the massless vortex-string chiral mode — has been theoretically extended to condensed matter systems(*24-27*) and photonic metamaterials (*28*), leading to a promising avenue to realise helical Majorana fermions(*29*) and topological fibres (*30*). However, experimental observation of such massless vortex-string chiral-anomaly modes remains elusive.

Here, we experimentally observe the vortex-string chiral modes in metamaterials. Specifically, we apply an inhomogeneous perturbation on the Yang-monopole phononic metamaterial. Such configurations can support a pair of vortex-string chiral modes with linear gapless dispersion and chiral anomaly propagation behaviour. Using pump-probe acoustic measurement, we directly observe the chiral anomaly flow of acoustic waves, the linear gapless dispersion, and the tightly-confined modal profiles along the string. Our findings not only provide a conclusive experimental advance on the observation of vortex-string chiral modes but also open the possibility of constructing high-performance and robust fibres for future information technologies.

As shown in Fig. 1a, we begin with a pair of Weyl monopoles with opposite first Chern numbers $C_1 = \pm 1$. The presence of a Higgs field $\Delta = p_1 + jp_2$, which is a complex scalar field, can mix the Weyl monopoles and delegate a nonzero mass to them when $\Delta$ is nonzero (27). By considering $(p_1, p_2)$ as two synthetic momentum parameters, the Weyl monopole pair can be upgraded to a Yang monopole (3) living in a five-dimensional (5D) momentum space defined by $\vec{k} = (k_x, k_y, k_z, p_1, p_2)$, where $(k_x, k_y, k_z)$ describes the conventional momentum vector. Yang monopole is described by a generalised Dirac Hamiltonian: $\mathbf{H} = \omega_Y \mathbf{I} + v_D(\vec{k} - \vec{k}_Y) \cdot \vec{\Gamma}$, where $\mathbf{I}$ is a 4×4 unitary matrix, $\vec{\Gamma} = (\Gamma_1, \Gamma_2, \Gamma_3, \Gamma_4, \Gamma_5)$ represents five 4×4 Gamma matrices satisfying the anticommutation relation $\{\Gamma_i, \Gamma_j\} = 2\delta_{ij}$, $\omega_Y$ is the energy of the Yang monopole, and $\vec{k}_Y = (k_x^Y, k_y^Y, k_z^Y, 0, 0)$ describes its location in 5D space. By integrating the curvature of the non-Abelian gauge field throughout the four-dimensional (4D) manifold enclosing the Yang monopole (shown in Fig. 1a), one can obtain a nonzero second Chern number $|C_2^Y| = 1$ (see Supplementary Information). As we will discuss in the following discussion, the vortex-string chiral mode is determined by the second Chern number of the Yang monopole.

In cosmology, cosmic strings emerge at spacetime positions where the Higgs field frustrates along a line. Similarly, a vortex string can be implemented in metamaterials when the complex scalar field $\Delta$ exhibits nonzero vorticity along a one-dimensional (1D) string extending in the spatial domain. As shown in Fig. 1, this can be implemented by mapping the spatial domain $\mathbf{r} = (x, y, z)$ to the two-dimensional (2D) parameter subspace,

$$\begin{bmatrix} p_1(\mathbf{r}) \\ p_2(\mathbf{r}) \end{bmatrix} = \Delta_0^{\max} \cdot \tanh(R/R_0) \cdot \begin{bmatrix} \cos(n \cdot \phi) \\ \sin(n \cdot \phi) \end{bmatrix}. \tag{1}$$

Here, $(R, \phi, z) = \left[\sqrt{x^2 + y^2}, \tan^{-1}(x/y), z\right]$ is the cylinder coordinates of the real space, and $R_0$ and $n$ represent the radius and vorticity of the vortex string, respectively. For arbitrary values of $R_0$ and $n$, the vortex-string topological chiral modes can be obtained analytically (see Supplementary Information). There always exist $n$-fold degenerate topological chiral modes protected by the second Chern number $C_2 = nC_2^Y$ defined in the 4D space $(k_x, k_y, k_z, \phi)$, where $C_2^Y$ is the second Chern number of a Yang monopole describing the non-Abelian topological charge surrounded by the 4D cylindrical manifold $(k_x,$ $k_y, k_z, \theta$) shown in Fig. 1a. It can be theoretically calculated that each Yang monopole can lead to $n$ degenerate topological zero modes confined near the core of the vortex string. Without losing any generality, we study the case of $R_0 \to 0$ and $n = 1$ in our experiments, so that each Yang monopole supports a single topological mode at the core of the vortex string with a chiral wavefunction and a linear dispersion along $z$-direction (see Figs. 1b and c):

$$\psi(k_z, R, z) = e^{jk_z z} e^{-R \cdot \Delta_0^{max}/v_D} \begin{pmatrix} e^{j\pi/4} \\ 0 \\ 0 \\ e^{-j\pi/4} \end{pmatrix}, \quad \omega(k_z) = \omega_Y + v_D \cdot k_z. \qquad (2)$$

To experimentally realise the vortex-string chiral modes, we propose a realistic design of Yang-monopole phononic metamaterials (see Fig. S1). The phononic metamaterial exhibits a hexagonal lattice in the $x$-$y$ plane with a lattice constant of $a_0 = 41$ mm. Each unit cell contains two substructures A ($0 < z < h/2$) and B ($h/2 < z < h$), with exactly the same geometry so that the entire structure has a half-lattice translational symmetry along the $z$ direction. Here the lattice constant in $z$-direction is $h = 32$ mm. Each substructure contains six spiral air channels connecting the plane air layers with fan-shaped scatters. The spiral air channels have a twist angle $\zeta = 92.9°$, which has been elaborately optimised so that the structure supports an ideal Dirac monopole (i.e. a pair of iso-frequency Weyl monopoles with opposite charges) at K/K' point in Fig. 1d. This Dirac monopole can be further promoted to Yang monopoles in 5D space if the substructures A and B (marked in red and blue in Fig. S1d) have different geometries defined by parameters $(p_1, p_2)$, where $p_1$ shrinks (expands) the size of the spiral air channels to $\alpha_0 \mp \alpha(p_1)$ and $w_0 \mp w(p_1)$ with $\alpha(p_1) = p_1 \cdot 7.5°$ and $w(p_1) = p_1 \cdot 2.375$ mm, and $p_2$ rotates the fan-shape scatters counterclockwise (clockwise) by $\beta(p_2) = \text{asin}(p_2)/3$ in radian unit. The numerically calculated projected band diagrams in momentum space with $(p_1, p_2) = (0, 0)$ (see Fig. 1d) and in parameter space with $(k_x, k_y, k_z) = (\pm 4\pi/3a_0, 0, 0)$ (see Figs. 1e and f) confirm that the proposed phononic metamaterial can support a pair of Yang monopoles at $\vec{k}_Y = (k_x^Y, k_y^Y, k_z^Y, 0, 0) = (\pm 4\pi/3a_0, 0, 0, 0, 0)$. Due to the time-reversal invariance of the system, the two Yang monopoles located at the K and K' points naturally exhibit opposite second Chern numbers $C_2^Y = \pm 1$ (see Supplementary Information).

The experimentally realised phononic metamaterial device supporting vortex-string chiral modes is shown in Figure 2. The device is the opposite structure of the air region shown in Fig. 1 and is fabricated by a three-dimensional (3D) printing technique with photopolymer materials. The full dimension of the device is 426 × 492 × 576 mm, containing 18 unit cells in the $z$ direction. Vortex string is implemented by mapping the spatial cylinder coordinates $(R, \phi, z)$ to the synthetic parameter space $(p_1, p_2)$ using Eq. (1), so that each unit cells exhibit different values of complex scalar field $\Delta = p_1 + jp_2$ (see Figs. 2d), and the entire phononic metamaterial is an inhomogeneous structure in the *x-y* plane. During the experiment, the acoustic source is threaded into the bottom or top layer of the device through a round hole near the core of vortex strings (marked by the arrow in Fig. 2c). Figure 2e shows the measured spatial distribution of normalised acoustic pressure along *z* direction when the source is placed at the bottom or top layer, suggesting that the acoustic waves can propagate along the core of vortex string with wavevector around $k_z = 0$ at a frequency of 4.13 kHz, which coincides with the simulated frequency of Yang monopole $\omega_Y/2\pi = 4.13$ kHz in Fig. 1d. Figure 2f shows the simulated energy band diagram of the device, which agrees well with the experimentally measured dispersions (Figs. 2g and h), suggesting that such vortex-string phononic metamaterial can support two counterpropagating massless topological modes in the bulk bandgap frequency region. With the experimentally measured acoustic pressure, we have also calculated the loss and group velocity of the vortex-string chiral modes, which also agree well with numerically calculated values (see Supporting Information). These results confirm that there are two counterpropagating modes along the vortex strings with massless linear dispersions. Note that the coupling between these two modes is negligible in the present scale both numerically and experimentally in Figs. 2f–h. As will be shown in the following study, these two counterpropagating modes originate from the two Yang monopoles located at K and K' points, respectively.

Chiral anomaly propagation is a key feature of the vortex-string chiral modes. Figures 3a and b illustrate the simulated modal profiles at $k_z = 0.1 \times 2\pi/h$, where the eigenfrequencies of the forward and backward propagating vortex-string topological modes are $\omega(k_z) = 4.246$ kHz and $\omega(k_z) = 4.020$ kHz, respectively. Figures 4c and d (Figures 4e and f) show the modal profile of forward (backward)-propagating topological mode measured on the third layer along positive (negative) *z* direction at a frequency of 4.246 kHz (4.020 kHz). The simulated and measured modal profiles confirm that the topological modes are tightly confined to near the core of the vortex strings, which agrees with the theoretically calculated modal profiles in Eq. (2). Most importantly, we project the measured acoustic-wave signals into momentum space by conducting spatial Fourier transform and find that the acoustic intensity of forward (backward) propagating vortex-string topological mode is confined around the K' (K) point in the momentum space in the bulk bandgap frequency region. This suggests that propagating directions of the vortex-string topological modes are strictly locked to the non-Abelian topological charge of Yang monopoles. Based on the measured results, we analyse the valley polarisation of the forward and backward acoustic waves along the vortex string (see Fig. 3k), which is defined by $(I_{K'} - I_K)/(I_{K'} + I_K)$, where $I_{K'}$ and $I_K$ represents the intensity of acoustic waves at K' and K points, respectively. Figure 3k clearly suggests that the forward (backward) propagating acoustic waves are mainly K' (K) polarised in the bulk bandgap region (grey shaded region in Fig. 3k). Based on the experimental results, we have also analysed the four-component wavefunction $\psi = (\psi_1, 0, 0, \psi_4)$ by plotting the amplitude and phase of $\psi_4/\psi_1$ in Fig. 3l. According to Eq. (2), $\psi_4/\psi_1$ equals $\exp(\mp i\pi/2)$ for forward and backward propagating mode. The measured values of $\psi_4/\psi_1$ in Fig. 3l slightly deviate in the amplitude, but the phase agrees well with the theoretical prediction. All these results provide conclusive evidence of chiral anomaly propagating behaviours of acoustic waves along the vortex strings.

In cosmology, the massless linear dispersion and chiral anomaly of the fermionic mode play an important role in making the vortex string "superconducting". These features shown in Figs. 2 and 3 are also unique and important in the realm of classical-wave metamaterial systems because they make the vortex string a promising candidate towards practical topological fibres or topological coaxial cables. One notes that these features (massless linear dispersion and chiral anomaly) are not available if the 3D metamaterial system is reduced to 2D and the 1D vortex-string chiral mode is reduced to 0D Dirac-vortex mode(*31-34*). Besides, our design is also different from the previous theoretical proposals of topological fibres in magnetic gyroid photonic metamaterials(*30*) because the vortex-string chiral modes in our systems can propagate without the necessity of breaking time-reversal symmetry, as far as the two Yang monopoles do not mix with each other.

In conclusion, we experimentally construct a vortex string in an inhomogeneous phononic metamaterial and directly observe the vortex-string topological chiral modes with massless dispersions and chiral anomaly. Our work not only reveals that phononic metamaterials are a useful tool to emulate various cosmic topological phenomena in astrophysics, but also provides a practical strategy for realising topological fibre in the classical-wave systems, finding wide applications in enhancing the robustness of signal processing and communication systems.

# References

1. H. Weyl, Electron and gravitation. *Z. Phys.* **56**, 330–352 (1929).
2. P. A. M. Dirac, Quantised singularities in the electromagnetic field. *Proc. R. Soc. London A Math. Phys. Eng. Sci.* **133**, 60–72 (1931).
3. C. N. Yang, Generalization of Dirac's monopole to SU2 gauge fields. *J. Math. Phys.* **19**, 320–328 (1978).
4. N. R. Cooper, J. Dalibard, I. B. Spielman, Topological bands for ultracold atoms. *Rev. Mod. Phys.* **91**, 015005 (2019).
5. M. Z. Hasan, C. L. Kane, Colloquium: Topological insulators. *Rev. Mod. Phys.* **82**, 3045–3067 (2010).
6. L. Lu, J. D. Joannopoulos, M. Soljačić, Topological photonics. *Nat. Photon.* **8**, 821–829 (2014).
7. H. Xue, Y. Yang, B. Zhang, Topological acoustics. *Nat. Rev. Mater.* **7**, 974–990 (2022).
8. S.-Y. Xu *et al.*, Discovery of a Weyl fermion semimetal and topological Fermi arcs. *Science* **349**, 613–617 (2015).
9. L. Lu *et al.*, Experimental observation of Weyl points. *Science* **349**, 622–624 (2015).
10. B. Yang *et al.*, Ideal Weyl points and helicoid surface states in artificial photonic crystal structures. *Science* **359**, 1013–1016 (2018).
11. C. He *et al.*, Acoustic analogues of three-dimensional topological insulators. *Nat. Commun.* **11**, 2318 (2020).
12. F. Li, X. Huang, J. Lu, J. Ma, Z. Liu, Weyl points and Fermi arcs in a chiral phononic crystal. *Nat. Phys.* **14**, 30–34 (2018).
13. V. Peri, M. Serra-Garcia, R. Ilan, S. D. Huber, Axial-field-induced chiral channels in an acoustic Weyl system. *Nat. Phys.* **15**, 357–361 (2019).
14. L.-Y. Zheng, J. Christensen, Dirac hierarchy in acoustic topological insulators. *Phys. Rev. Lett.* **127**, 156401 (2021).
15. S. Sugawa, F. Salces-Carcoba, A. R. Perry, Y. Yue, I. B. Spielman, Second Chern number of a quantum-simulated non-Abelian Yang monopole. *Science* **360**, 1429–1434 (2018).
16. S. Ma *et al.*, Linked Weyl surfaces and Weyl arcs in photonic metamaterials. *Science* **373**, 572–576 (2021).
17. H. B. Nielsen, P. Olesen, Vortex-line models for dual strings. *Nucl. Phys. B* **61**, 45–61 (1973).
18. T. W. Kibble, Topology of cosmic domains and strings. *J. Phys. A* **9**, 1387–1398 (1976).
19. A. A. Abrikosov, The magnetic properties of superconductiang alloys. *J. Phys. Chem. Solids* **2**, 199–208 (1957).
20. C. Bäuerle, Y. M. Bunkov, S. N. Fisher, H. Godfrin, G. R. Pickett, Laboratory simulation of cosmic string formation in the early universe using superfluid $^3$He. *Nature* **382**, 332–334 (1996).
21. I. Chuang, R. Durrer, N. Turok, B. Yurke, Cosmology in the laboratory: Defect dynamics in liquid crystals. *Science* **251**, 1336–1342 (1991).
22. E. Witten, Superconducting strings. *Nucl. Phys. B* **249**, 557–592 (1985).
23. E. M. Chudnovsky, G. B. Field, D. N. Spergel, A. Vilenkin, Superconducting cosmic strings. *Phys. Rev. D* **34**, 944–950 (1986).
24. A. A. Zyuzin, A. A. Burkov, Topological response in Weyl semimetals and the chiral anomaly. *Phys. Rev. B* **86**, 115133 (2012).
25. Z. Wang, S.-C. Zhang, Chiral anomaly, charge density waves, and axion strings from Weyl semimetals. *Phys. Rev. B* **87**, 161107 (2013).
26. C.-X. Liu, P. Ye, X.-L. Qi, Chiral gauge field and axial anomaly in a Weyl semimetal. *Phys. Rev. B* **87**, 235306 (2013).
27. T. Schuster, T. Iadecola, C. Chamon, R. Jackiw, S.-Y. Pi, Dissipationless conductance in a topological coaxial cable. *Phys. Rev. B* **94**, 115110 (2016).
28. R. Bi, Z. Wang, Unidirectional transport in electronic and photonic Weyl materials by Dirac mass engineering. *Phys. Rev. B* **92**, 241109 (2015).

29. E. J. König, P. Coleman, Crystalline-symmetry-protected helical Majorana modes in the iron pnictides. *Phys. Rev. Lett.* **122**, 207001 (2019).
30. L. Lu, H. Gao, Z. Wang, Topological one-way fiber of second Chern number. *Nat. Commun.* **9**, 5384 (2018).
31. X. Gao *et al.*, Dirac-vortex topological cavities. *Nat. Nanotechnol.* **15**, 1012–1018 (2020).
32. P. Gao *et al.*, Majorana-like zero modes in Kekulé distorted sonic lattices. *Phys. Rev. Lett.* **123**, 196601 (2019).
33. J. Noh *et al.*, Braiding photonic topological zero modes. *Nat. Phys.* **16**, 989–993 (2020).
34. C. Sheng *et al.*, Bound vortex light in an emulated topological defect in photonic lattices. *Light Sci. Appl.* **11**, 243 (2022).

**Materials and Methods**

**Simulation.** We used commercial software (COMSOL Multiphysics) to numerically calculate the band structures and modal profiles, with an air density 1.18 kg/m$^3$ and a sound speed 343 cm$^{-1}$. During the simulation, air-solid interfaces are assumed to be hard acoustic boundaries. For the calculation of bulk band diagrams in Fig. 1, we used Bloch periodic boundary conditions in all three directions. For the calculation of vortex-string chiral modes in Figs. 2 and 3, we applied Bloch periodic boundary conditions only in the $z$ direction, and plane-wave radiation boundary conditions in the $x$ and $y$ directions.

**Experiments.** The experimental samples were fabricated from photosensitive resin via stereolithographic 3D printing. For the dispersion measurements in Fig. 2, a broadband acoustic signal is launched from a balanced armature speaker of around a 1 mm radius, driven by a power amplifier, and located near the core of vortex strings at the top or bottom layer of the sample. Each acoustic probe is a microphone (Brüel & Kjær Type 4961, of about 3.2 mm radius) in a sealed sleeve with a tube of 1 mm radius and 250 mm length. The probes can be threaded into the sample along the horizontal air regions to scan different positions within each layer of the sample. The measured data was processed by a Brüel & Kjær 3160-A-022 module to extract the frequency spectrum, with 2 Hz resolution. Spatial Fourier transforms are applied to the complex acoustic pressure signals to obtain the dispersion relation and field distributions.

**Funding:**

The work is sponsored by the Key Research and Development Program of the Ministry of Science and Technology under Grants 2022YFA1405200 (Y.Y.), No. 2022YFA1404704 (H.C.), 2022YFA1404902 (H.C.), and 2022YFA1404900 (Y.Y.), the National Natural Science Foundation of China (NNSFC) under Grants No. 62175215 (Y.Y.), and No. 61975176 (H.C.), the Key Research and Development Program of Zhejiang Province under Grant No.2022C01036 (H.C.), the Fundamental Research Funds for the Central Universities (2021FZZX001-19) (Y.Y.), the Excellent Young Scientists Fund Program (Overseas) of China (Y.Y.), the National Natural Science Foundation of China under Grant Nos. 12274183 (H.X.S.) and 12174159 (Y.G.), the National Key Research and Development Program of China under Grant No. 2020YFC1512403 (S.Q.Y.), and Hong Kong Research Grants Council under Grant No. N_HKU750/22 (X.Z.).

**Competing interests:** Authors declare that they have no competing interests.

**Data and materials availability:** All data are available in the main text or the supplementary materials.

Supplementary Materials
Figs. S1 to S2

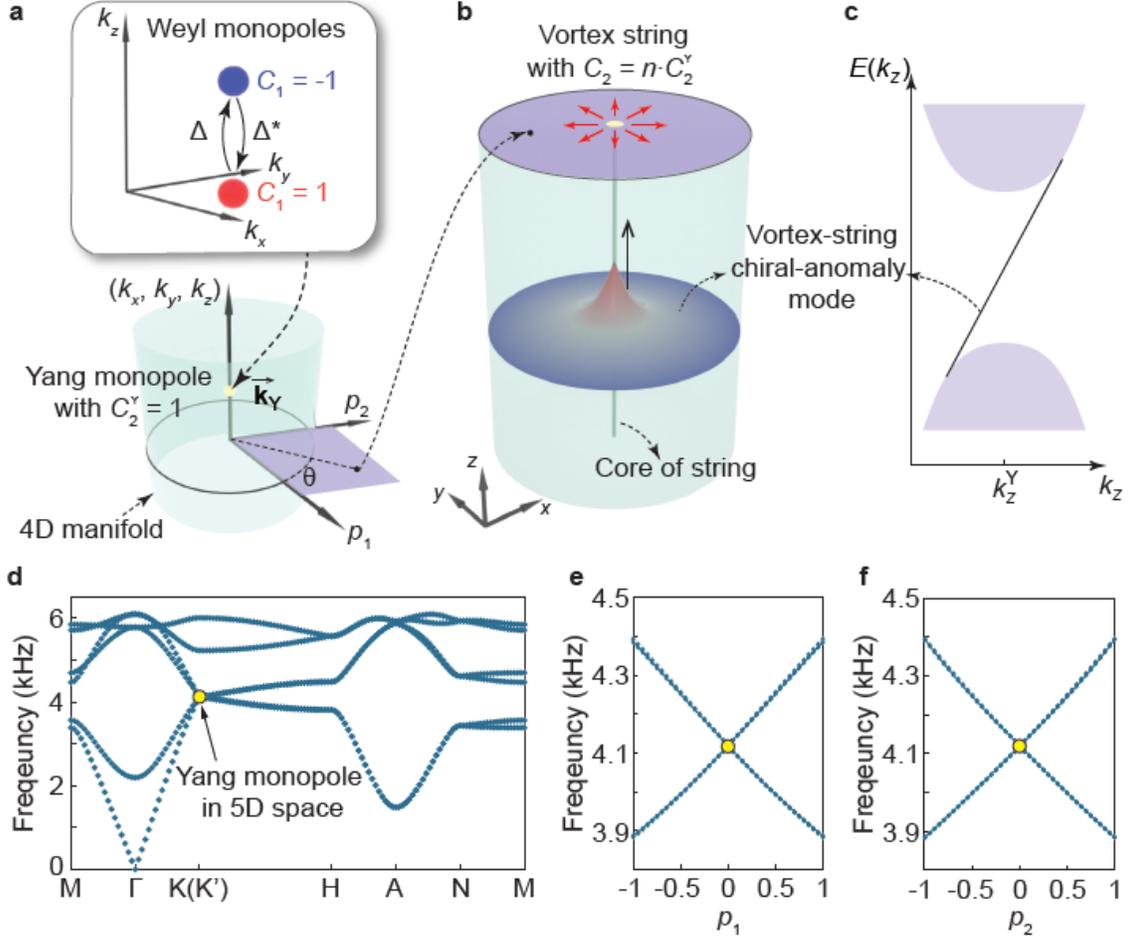

**Fig. 1 | Phononic metamaterials for realising vortex-string chiral modes. a,** Conceptual illustration of a pair of Weyl monopoles with opposite first Chern number. If Weyl monopoles are mixed to each other through a complex scalar field $\Delta = p_1 + jp_2$, it can be considered as a Yang monopole living in 5D space spanned by 3D momentum space $(k_x, k_y, k_z)$ and 2D parameter space $(p_1, p_2)$. **b,** Conceptual illustration of a vortex string. A vortex string is formed when the scalar field $\Delta = p_1 + jp_2$ exhibits nonzero vorticity $n$ around a 1D line in real space. This can be realised by properly defining a mapping from the real space to 2D parameter space $(p_1, p_2)$. Consequently, vortex-string chiral-anomaly modes emerge at the core of the strings. **c,** Massless linear dispersion of the vortex-string chiral-anomaly modes. This vortex-string chiral-anomaly mode is topologically protected by the nonzero second Chern number $C_2$, which originates from the non-abelian topological charge of Yang monopoles surrounded by the 4D manifold shown in **a**. **d,** Numerically calculated band diagrams in momentum space with $(p_1, p_2) = (0, 0)$. **e, f,** Numerically calculated dispersions along parameter axis $p_1$ (**e**) and $p_2$ (**f**) with $(k_x, k_y, k_z) = (\pm 4\pi/3a_0, 0, 0)$.

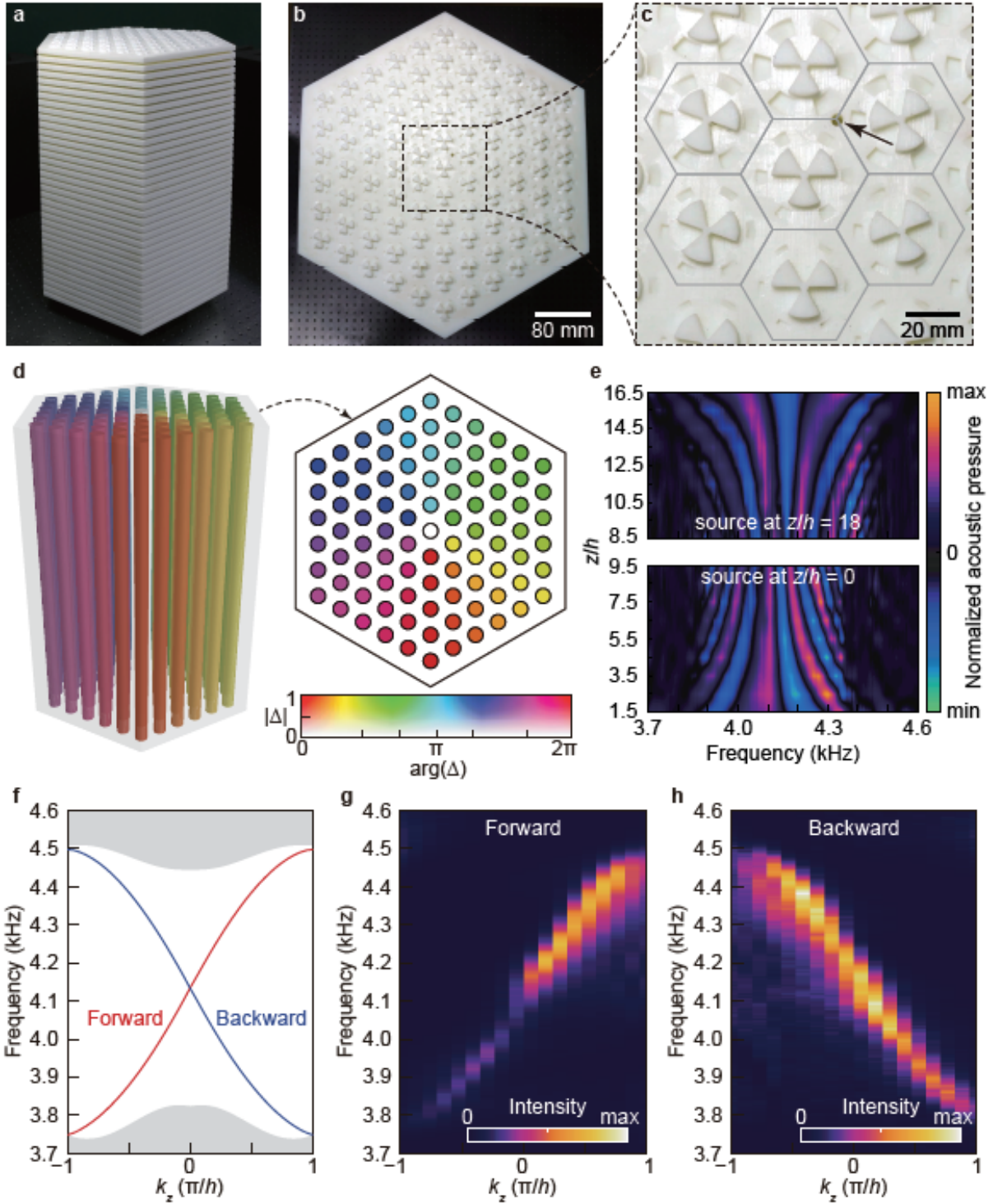

**Fig. 2 | Experimental observation of vortex string and its chiral modes in phononic metamaterials. a,** Oblique view of the 3D printed phononic metamaterial supporting vortex strings. **b,** Top view of the experimental device. The entire phononic metamaterial is an inhomogeneous structure in the *x-y* plane. **c,** Zoomed-in top-view picture of the structure near the core of the vortex string. A round circle is drilled in

the top layer to facilitate efficient excitation of acoustic waves in experiments. **d,** Spatial distribution of the complex scalar field $\Delta = p_1 + jp_2$ in the *x-y* plane. The phase of $\Delta$ experience a winding of $2\pi$ around the center of the device. **e,** Measured spatial distribution of acoustic pressure along *z* direction when the acoustic source is placed at the top ($z/h = 18$) and bottom ($z/h = 0$) layer respectively. **f.** Simulated energy band diagram of the fabricated device. There exists a pair of counterpropagating modes with linear dispersions in the bulk bandgap region. **g, h,** Experimentally measured dispersion when the acoustic waves propagate in the forward (**g**) and backward (**h**) direction.

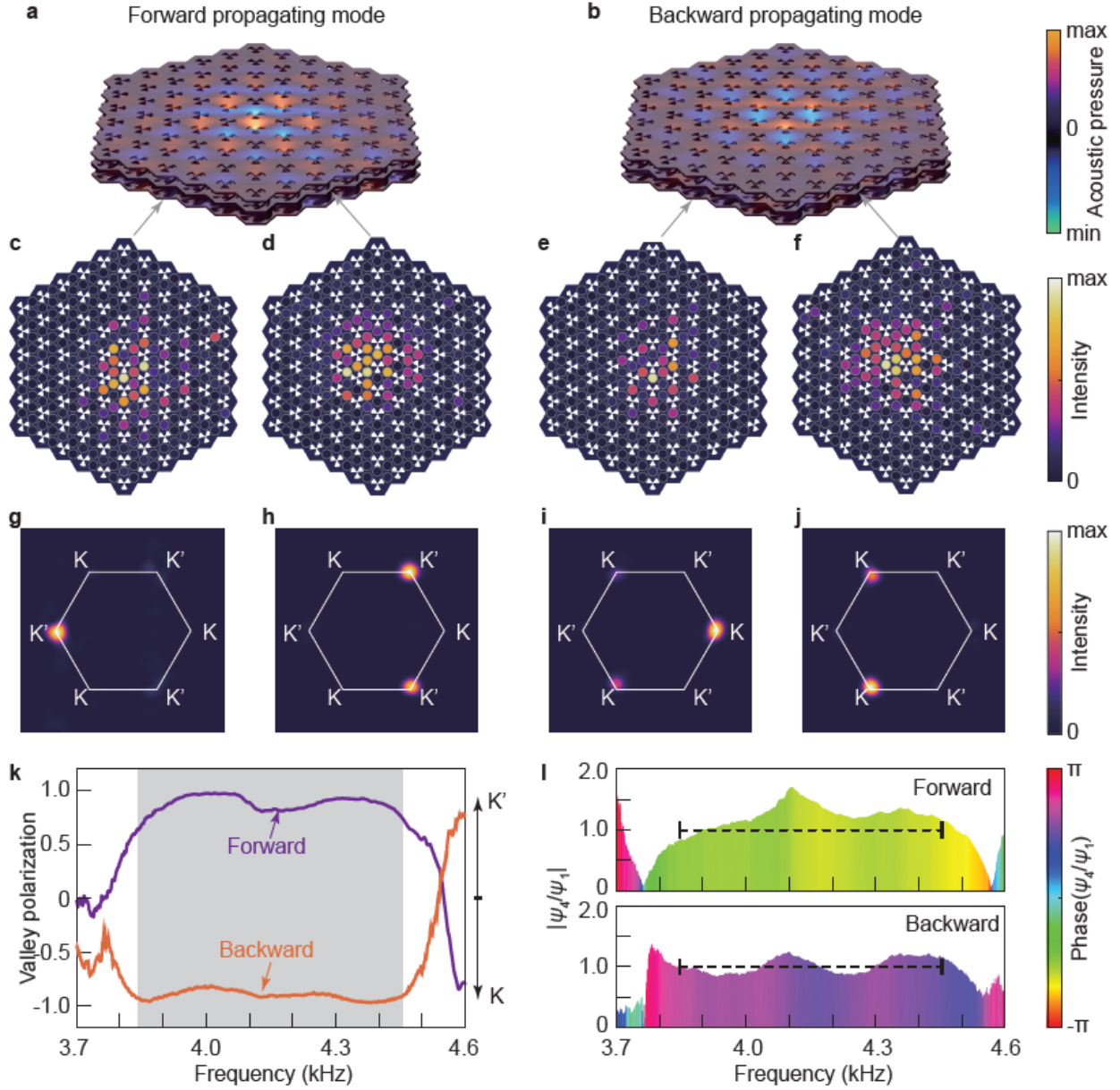

**Fig. 3 | Chiral anomaly behaviors of the vortex-string topological chiral modes. a, b,** Simulated modal profiles at $k_z = 0.1 \times 2\pi/h$, where the eigenfrequencies of the forward (**a**) and backward (**b**) propagating vortex-string topological modes are $\omega(k_z) = 4.246$ kHz and $\omega(k_z) = 4.020$ kHz, respectively. **c, d,** Modal profile of forward-propagating topological mode measured on the third layer along positive $z$ direction at frequency of 4.246 kHz. **e, f,** Modal profile of backward-propagating topological mode measured on the sixteenth layer along positive $z$ direction at frequency of 4.020 kHz. **g-i,** Momentum-space distribution of acoustic energies according to **c-f**, which suggests that propagating directions of the vortex-string topological modes are strictly locked to the K and K' points in the momentum space. **k,** Measured valley polarisation of the forward and backward propagating acoustic waves. In the bulk bandgap region (gray

shaded region), the intensity of acoustic waves is mostly near K' (K) point for forward (backward) propagating modes. **l**, Measured values of $\psi_4/\psi_1$ suggesting the chirality of the forward and backward modes. According to Eq. (2), it ideally takes the value of $\exp(\mp i\pi/2)$. In the bulk bandgap region (marked by dashed line), the amplitude of $\psi_4/\psi_1$ slightly deviate from 1 due to experimental error, but the phase of $\psi_4/\psi_1$ is around $\mp\pi/2$ for forward and backward propagating modes, which agree well with the theoretical prediction.